\documentclass[a4paper]{article}

\usepackage{INTERSPEECH2020}
\usepackage{multirow}
\usepackage{hyperref}

\title{Atss-Net: Target Speaker Separation via Attention-based Neural Network}
\name{Tingle Li$^{1,3}$, Qingjian Lin$^1$, Yuanyuan Bao$^1$, Ming Li$^{1,2}$}
\address{
  $^1$Data Science Research Center, Duke Kunshan University, Kunshan, China\\
  $^2$School of Computer Science, Wuhan University, Wuhan, China\\
  $^3$School of Computer Science and Technology, Tiangong University, Tianjin, China}
\email{ming.li369@dukekunshan.edu.cn}

\begin{document}

\maketitle
\begin{abstract}
Recently, Convolutional Neural Network (CNN) and Long short-term memory (LSTM) based models have been introduced to deep learning-based target speaker separation. In this paper, we propose an Attention-based neural network (Atss-Net) in the spectrogram domain for the task. It allows the network to compute the correlation between each feature parallelly, and using shallower layers to extract more features, compared with the CNN-LSTM architecture. Experimental results show that our Atss-Net yields better performance than the VoiceFilter, although it only contains half of the parameters. Furthermore, our proposed model also demonstrates promising performance in speech enhancement.
\end{abstract}

\section{Introduction}

We humans have the ability to focus on a target voice in the room where the noises are all around \cite{cherry1953some}, namely ``cocktail party effect", but machines have great difficulties in this task. However, this capability is essential in many back-end application, e.g. speaker verification \cite{rao2019target}, speaker diarization \cite{sell2018diarization} and speech recognition \cite{li2015robust, watanabe2017new, xiao2016study}.

Recently, a technique called blind speech separation was proposed to solve the aforementioned problem. Given a mixed utterance, this technique can automatically separate it into several different clean utterances that make it up. Specifically, the current DNN-based blind speech separation methods, such as Time-domain Audio Source Separation (Tas-Net) \cite{luo2018tasnet,luo2019conv,luo2018real}, Dual-path RNN (DPRNN) \cite{luo2019dual}, which directly model in the time-domain, have achieved good separation performance. However, these methods require the number of speakers to be known in advance, making it difficult to be applied in real world applications. In order to aggregate the exact number of speaker clusters using the K-Means algorithm, such methods like Deep clustering (DPCL) \cite{hershey2016deep, isik2016single} and Deep Attractor Network (DANet) \cite{chen2017deep}, need to be given the number of speakers in advance while in the inference phase.

Target speaker separation is one of the methods that address the above problem \cite{rao2019target, wang2018voicefilter}. Given a reference utterance of the target speaker, and a mixed utterance containing the target speaker, the target speaker separation system aims at filtering out the target speaker's voice from the mixed utterance. This technique is very useful when the back-end tasks only need the utterance of the target speaker. For example, in the keyword spotting task, supposing the user wakes up the devices in a restaurant, this technique can easily separate the speaker from background babble noise, if the speaker's voice is recorded in advance.

How to utilize the reference speech of the target speaker to handle the mixed signals is still a very challenging task. Recent successful studies \cite{rao2019target, wang2018voicefilter} extract speaker embeddings like i-vector \cite{garcia2011analysis} from the target utterance and concatenate with acoustic features in each frame. And then feed it into an LSTM layer together with the mixed spectrogram that passing through CNNs. However, this method has some limitations. First, it is difficult for LSTM to perform parallelized computation. Second, deeper CNN layers are used to acquire a wider receptive field, which means the number of model parameters is large.

The attention mechanism \cite{bahdanau2014neural} is a recent advance in neural network modeling. It learns one coefficient for every feature, which enables feature interactions that contribute differently to the final prediction. Besides, the importance of the feature interaction is automatically learned from data without any human domain knowledge. Recently, the attention-based methods \cite{shi2020furca, liu2020voice} have been proposed for the source separation task, which outperforms most of the state-of-the-art methods. Experimental results in \cite{liu2020voice} show that the attention mechanism is suitable for processing spectrogram, because it can be easily computed parallelly and solves long-term dependency problems well. Also, the attention mechanism allows us to use shallower CNNs for even better separation performance, compared with the CNN-LSTM architecture.

In this work, we try to explore the attention mechanism in the target speaker separation task, and propose an attention-based neural network model named Atss-Net. Experimental results show that Atss-Net has achieved a new state-of-the-art result, on the same dataset as in \cite{wang2018voicefilter}. Moreover, in order to evaluate the potential speech enhancement performance of our Atss-Net, we also added noise data such as pure music, songs (contain the voice of singers), TV noise, and so on.

\section{Target Speaker Separation}

\begin{figure}[htb]
\centering
\includegraphics[width=\linewidth]{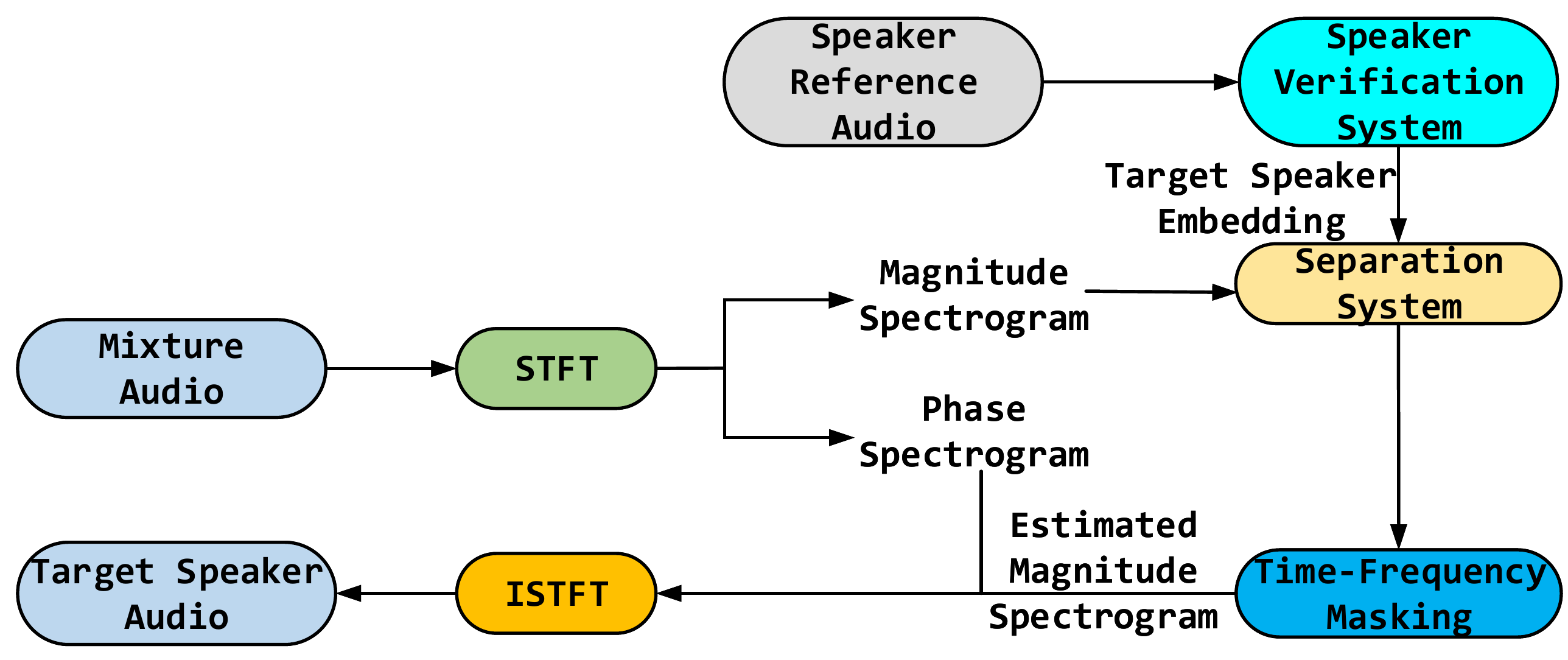}
\caption{The flow chart of our separation system.}
\label{fig:source_separation}
\end{figure}

The target speaker separation system consists of two parts: (1) speaker verification system, where we extract the deep speaker embedding of the target speaker from the reference speech signal; (2) speaker separation system, where we input the target speaker's embedding and the mixed spectrogram to generate the separated target speaker's spectrogram from the mixed one. These two systems are trained separately. The flow chart is shown in Figure \ref{fig:source_separation}.

\subsection{Speaker Verification System}

In our system, text-independent speaker verification serves as a discriminator, which judges which speaker's utterance should be separated out from the mixed utterance. Given a clean reference utterance of the target speaker, we extract 64-dimensional log-Mel-filterbank features (Fbank) with 25 ms frame length and 10 ms frameshift and use a frame-level energy-based voice activity detector (VAD) to filter out non-speech frames. Finally, we feed the features to a speaker verification system and get the target speaker embedding $\bm{e}_j \in \mathbb{R}^{1 \times \ddot{F}}$, where $\ddot{F}$ is the dimension of the frequency bins, and $j$ denotes the source of the target speaker.

\vspace{-0.25cm}
\subsection{Target Speaker Separation System}
Our network operates in the spectrogram domain as the baseline \cite{wang2018voicefilter} did. The mixed utterance can be defined as a linear combination of C speaker sources $s_1(t),...,s_c(t)$, where $s_i \in \mathbb{R}^{1 \times \widetilde{T}}$, and $\widetilde{T}$ is the duration of the utterance.
\vspace{-0.15cm}
\begin{equation}
x(t) = \sum_{i=1}^c s_i(t)
\label{eq1}
\end{equation}

\vspace{-0.15cm}
Then we perform the short-time Fourier transform (STFT) to get the spectrograms, so the sum of each source spectrogram $S_i(t,f)$ combines into mixed spectrogram $X(t,f) \in \mathbb{R}^{T \times \bar{F}}$, where $T$ and $\bar{F}$ are the dimension of time and frequency bin axes respectively.
\vspace{-0.4cm}
\begin{equation}
X(t, f) = \sum_{i=1}^c S_i(t,f)
\label{eq2}
\end{equation}

Besides, we need to input the target speaker embedding $\bm{e}_j \in \mathbb{R}^{1 \times \ddot{F}}$ to the model ($1 \!\leq\! j \!\leq\! c$). So we duplicate the speaker embedding for $T$ times to get the extended target speaker embedding $E_j(t,f) \in \mathbb{R}^{T \times \ddot{F}}$.

Similar to the baseline \cite{wang2018voicefilter}, we assume the phase of the mixture is the same as that of the separated utterance, so we only input the magnitude spectrogram $|X(t,f)|$ to the model together with the extended target speaker embedding $E_j(t,f)$ to estimate the time-frequency mask $M_j$.
\begin{equation}
M_{j}(t,f)=g(|X(t,f)|,E_j(t,f))
\label{eq4}
\end{equation}
where $g(t,f)$ is the target speaker separation system.

Therefore, the estimated magnitude spectrogram $|\widehat{X}_j(t, f)|$ is calculated by performing the element-wise product between the mixed magnitude spectrogram and the estimated time-frequency mask.
\begin{equation}
|\widehat{X}_j(t, f)|=|X(t, f)| \odot M_{j}(t, f))
\end{equation}
where $\odot$ denotes the element-wise product operation.

So this task can be denoted as minimizing the squared $l_2$-norm between the ground truth and the estimated magnitude:
\vspace{-0.15cm}
\begin{equation}
\mathcal{L}=\underset{M}{\arg \min } \left\||X_j(t, f)| - |\widehat{X}_j(t, f)| \right\|_{2}^{2}
\label{eq6}
\end{equation}

In addition, the reconstruction of time domain target speaker signal $\widehat{s}_j$ is accomplished by combining the estimated magnitude spectrogram and the mixed phase to perform inverse short-time Fourier transform (ISTFT), which is:
\begin{equation}
\widehat{s}_{j}(t)=\mathrm{ISTFT}(|\widehat{X}_j(t, f)| \times e^{\angle X(t, f)i})
\label{eq5}
\end{equation}
where $\angle X(t, f)$ is the phase of mixed utterance.

\section{The proposed Atss-Net}

\begin{figure}[htb]
\includegraphics[width=\columnwidth]{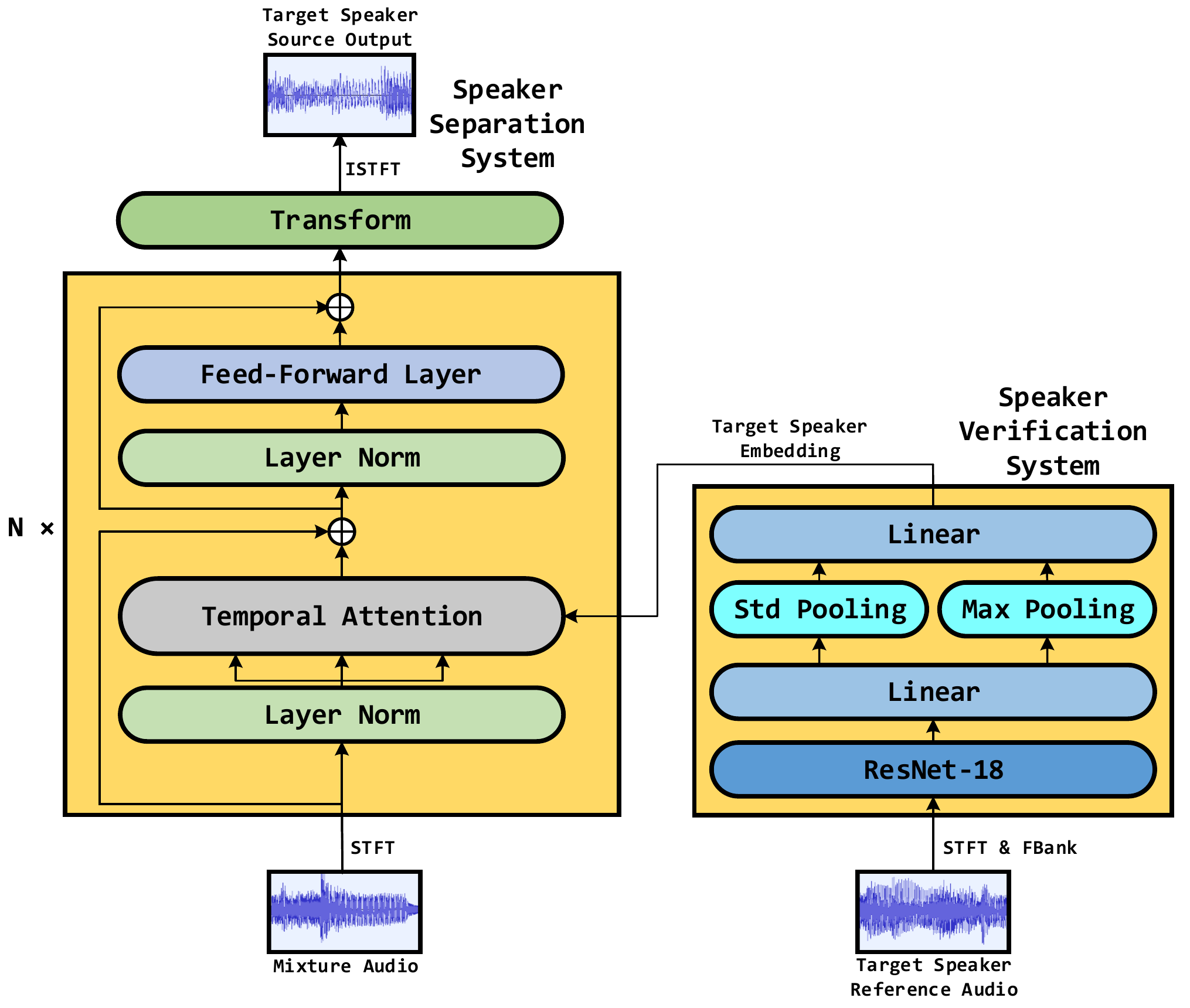}
\caption{Model architecture of the Atss-Net.}
\label{fig:Architecture}
\end{figure}

As shown in Figure \ref{fig:Architecture}, the speaker verification system in our Atss-Net is based on \cite{cai2020fly}. And the speaker separation system consists of $N$ Attention blocks and a Transform block. Specifically, each Attention block consists of a Multi-Head Temporal Scaled Dot-Product Attention layer (Section 3.2-3.4), a Feed-Forward layer (Section 3.5), and two Layer Normalization layers (Section 3.6). The Transform block includes a dense layer and a CNN layer with the kernel size $3 \times 3$, which aims to compress the channel dimension of the feature maps form $d_k$ to $1$ and generate the estimated mask.

\subsection{Deep ResNet-based Speaker Embedding}
Our deep speaker embedding extraction module includes a ResNet-18 front end \cite{he2016deep}, two parallelized pooling layers and two linear layers. First, the ResNet-18 module transforms L-frames Fbanks to feature maps, with the channel settings of \{16, 32, 64, 128\}, the kernel size of $3 \times 3$ and the padding size of 1. For each channel, the mean-pooling and standard-pooling are applied respectively, and the outputs are concatenated together. Then, we feed the output into two linear layers and the softmax function sequentially, generating speaker likelihoods. The size of the second linear layer equals the number of speakers in the training data. While in the inference phase, the second linear layer will be removed and the output of the first linear layer will be used as the target speaker embedding.

\subsection{Scaled Dot-Product Attention}
Self-Attention is a mechanism that can relate different time steps of each magnitude feature map to compute their representations. There are many types of attention mechanisms \cite{anderson2018bottom, oktay2018attention, vaswani2017attention}, and here we use the dot-product \cite{vaswani2017attention}. In our model, the magnitude feature maps are separately input to three different CNN layers with the kernel size $1 \times 1$, to get the query $Q$, keys $K$, and values $V$ respectively. Then, we compute the dot products of the query $Q$ with keys $K$, divide each by $ \sqrt{d_{k}} $, where $d_k$ is the channel dimension of the magnitude feature maps. Finally, a softmax function is applied to obtain the weights on the values V. The output of dot-product attention is computed as:
\begin{equation}
\mathrm{Attention}({Q},{K},{V}) = \mathrm{softmax}\left( {\frac{{{Q}{{K}^T}}}{{\sqrt {{d_k}} }}} \right){V}
\end{equation}
where $ 1 / \sqrt{d_{k}} $ is used to prevent softmax function into regions that have very small gradients.

\subsection{Multi-Head Attention}
Since the duration of each utterance is short, we need to find a way to make full use of sampling features and compute the attention value more precisely. Multi-Head Attention is a good option. Specifically, it performs Scaled Dot-Product Attention for $h$ times repeatedly, then concatenates all the results following the channel axis, and finally performs convolution to restore the origin shapes. The formula is as follows:
\begin{equation}
\begin{split}
\mathrm{MultiHead}({Q},{K},{V}) = \mathrm{Concat}(\mathrm{head}_1, \mathrm{head}_2, \\
\cdots, \mathrm{head}_i, \cdots, \mathrm{head}_h) \cdot W^O
\end{split}
\end{equation}
where $\mathrm{head}_i=\mathrm{Attention}(Q\!\cdot\!{W_i}^Q,K\!\cdot\!{W_i}^K,V\!\cdot\!{W_i}^V)$, $W^Q,W^K,W^V$ are there different CNN layers with the kernel size $1 \times 1$ and $W^O$ is a CNN layer with the kernel size $3 \times 3$.

\subsection{Temporal Attention}
\begin{figure}[tb]
\centering
\includegraphics[scale=0.5]{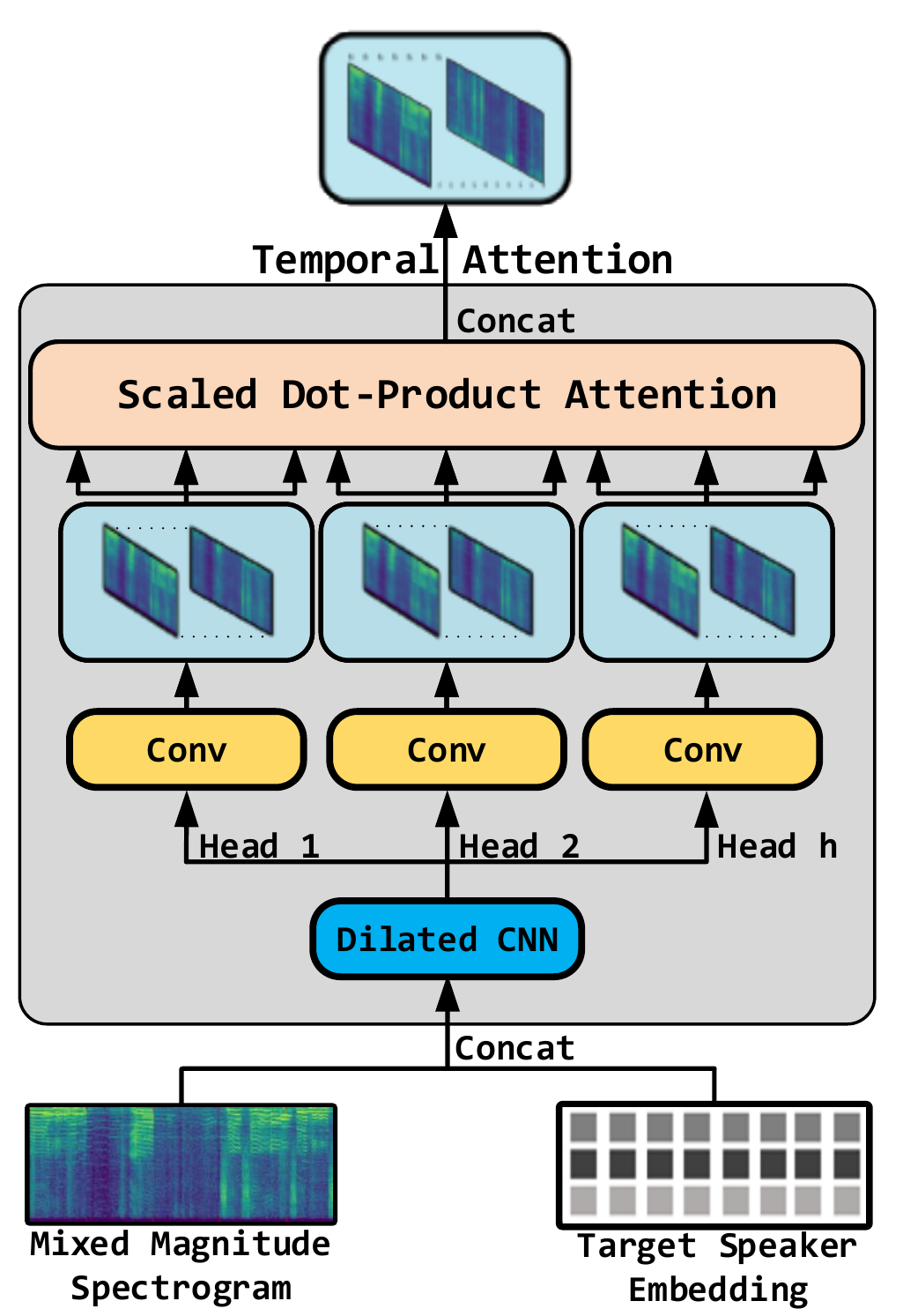}
\caption{Model architecture of the Temporal Attention.}
\label{fig:Temporal_Attention}
\end{figure}

The architecture of Temporal Attention (TA) is shown in Figure \ref{fig:Temporal_Attention}. Specifically, we first concatenate the mixed magnitude spectrogram $|X(t,f)| \in \mathbb{R}^{T \times \bar{F}}$and the extended target speaker embedding $E_j(t,f) \in \mathbb{R}^{T \times \ddot{F}}$ following the frequency bin axis to get the output $R(t,f) \in \mathbb{R}^{T \times F}$, and $F = \bar{F} + \ddot{F}$:
\begin{equation}
R(t,f) = \mathrm{Concat}(|X(t,f)|,E_j(t,f))
\end{equation}
where $|X(t,f)|$ and $E_j(t,f)$ are the magnitude of the mixed spectrogram and extended speaker embedding respectively.

Then, a group of Dilated CNN (DCNN) layers \cite{yu2015multi} is employed as the feature extractor, where details can be seen in Table 1. After that, Multi-Head Attention is performed on the frequency bin axis, and each head stores the complete temporal information. The formulas are as follows:
\begin{gather}
Q',K',V' = \mathrm{DCNN}(R(t,f)) \\
\mathrm{TA} = \mathrm{MultiHead}(Q',K',V')
\end{gather}

\begin{table}[tb]
\centering
\caption{Parameters of Dilated CNN.}
\begin{tabular}{cccc}
\toprule
\textbf{Layer} & \textbf{Kernel Size} & \textbf{Dilation} & \textbf{Output Channel} \\
\midrule
CNN 1 & $5 \times 5$ & $1 \times 1$ &  64\\
CNN 2 & $5 \times 5$ & $2 \times 1$ &  64\\
CNN 3 & $5 \times 5$ & $4 \times 1$ &  64\\
CNN 4 & $5 \times 5$ & $8 \times 1$ &  64\\
CNN 5 & $5 \times 5$ & $16 \times 1$ &  64\\
CNN 6 & $1 \times 1$ & $1 \times 1$ &  64\\
\bottomrule
\end{tabular}
\end{table}

\vspace{-0.5cm}
\subsection{Feed-Forward Layer}
Feed-Forward layer is another core module of our Atss-Net. It contains two linear transformations with a ReLU activation \cite{agarap2018deep} in between. The dimensions of input and output magnitude feature maps are $\mathbb{R}^ {d_k \times T \times F}$, and the dimension of the hidden layer is $\mathbb{R}^ {d_k \times T \times \widetilde{F}}$, where $\widetilde{F}$ is four times as much as $F$. The computing formula is as follows:
\vspace{-0.25cm}
\begin{equation}
\mathrm{FFN}(x)=\max \left(0, x W_{1}+b_{1}\right) W_{2}+b_{2}
\end{equation}
where x is the output of Temporal Attention (TA) module , $W_{1} \in \mathbb{R}^ {F \times \widetilde{F}}$, $W_{2} \in \mathbb{R}^ {\widetilde{F} \times F}$ and the biases $b_{1} \in \mathbb{R}^ {\widetilde{F}}$, $b_{2} \in \mathbb{R}^ {F}$.

\vspace{-0.25cm}
\subsection{Layer Norm}
Layer Normalization and residual connection are employed to each sub-block for effective training. So the output of each sublayer is:
\begin{equation}
x + \mathrm{Sublayer}(\mathrm{LayerNorm}(x))
\end{equation}
where $\mathrm{Sublayer}(x)$ could be the Temporal Attention module, and the Feed-Forwawrd layer module.

\vspace{-0.2cm}
\section{Experiment Setup}
\subsection{Data Description}
% \vspace{-0.15cm}
\subsubsection{Voxcelab Dataset}
Voxceleb \cite{nagrani2017voxceleb} is a large text-independent speaker verification dataset that includes more than 100,000 words from 1251 celebrities ``in the wild''. We use the official split training and validation set together as our development dataset. Specifically, there are 1211 celebrities in the training dataset, while the test dataset contains 4715 utterances from the rest 40 celebrities. Also, there are 37720 pairs of trials in total, where 18860 are true trials. And our system has a 2.97\% equal error rate (EER) on this test dataset.

\vspace{-0.2cm}
\subsubsection{Librispeech Dataset}
The Librispeech dataset \cite{panayotov2015librispeech} is used for the target speaker separation task as the baseline \cite{wang2018voicefilter} did, where the training set contains 2338 speakers, and the test set contains 73 speakers. Also, each utterance contains the voice of one speaker.

In detail, the training set and test set are generated on-the-fly as follows: (1) randomly select two speakers from the speaker list that \cite{wang2018voicefilter} provided; (2) for each speaker, randomly sample utterances with 3s. (3) use the FFmpeg\footnote{\url{https://github.com/FFmpeg/FFmpeg/}} to normalize the volume of selected utterances and mix them.

\vspace{-0.3cm}
\subsubsection{AISHELL-2 Dataset}
AISHELL-2 \cite{du2018aishell} is a Chinese Mandarin dataset that contains 1991 speakers and 87456 utterances in total, and the duration of each utterance is varied from 2s to 10s. And we use it for speech enhancement experiments here. Specifically, we randomly selected 1951 speakers and 85000 utterances for training, and the others for testing. When it comes to the background noise, we picked 50 pure music from the CCMixter dataset \cite{liutkus2015scalable}, 150 English songs from the MUSDB18 dataset \cite{rafii2017musdb18}, 25 Chinese songs and 200 far-field TV utterances from the Internet, in a total of 425 utterances.

The mixed utterances are simulated on-the-fly in 3 steps: (1) randomly select one speaker from the speaker list, and then randomly select two utterances of that speaker; utterance one is used as a referenced utterance, and utterance two is used to mix with background noise; (2) randomly select one utterance from the noise list, then randomly cut it to the same dimension as the selected speaker's utterance two; (3) mix these two utterances using the signal-to-noise ratio (SNR) uniformly sampled from 5dB to 20dB.

\vspace{-0.2cm}
\subsection{Training Setup}
We train the model using PyTorch \cite{paszke2019pytorch} with 4 NVIDIA TITAN RTX GPU. In order to compare our model with the baseline \cite{wang2018voicefilter}, the hyperparameters of our model are roughly consistent with \cite{wang2018voicefilter}. Specifically, the frame length and frame step of the STFT are set to 400 and 160 respectively and a 512-point discrete Fourier transform is performed. Also, the networks are trained to estimate the magnitude spectrogram using the Adam optimizer \cite{kingma2014adam}, with an initial learning rate of 0.0001. Also, due to the limitation of the GPU memory, we can only train Atss-Net with 3 Temporal Attention modules, 2 heads and 64 channel dimensions. Besides, we set the batch size to 16 and set the maximum epoch to 50. If the validation loss does not descend after 10 epoch, the early stop will be performed and the training phase will stop. Last but not least, we use signal-to-distortion ratio (SDR) \cite{vincent2006performance} as the evaluation metric for target speaker separation, and use the perceptual evaluation of speech quality (PESQ) ITU-T P.862 \cite{rix2001perceptual} to evaluate the speech enhancement performance. For both PESQ and SDR metrics, the higher number indicates the better performance.

\vspace{-0.3cm}
\section{Results}
\subsection{Target Speaker Separation Experiment}
Since the baseline \cite{wang2018voicefilter} is not open source, we try to build this system ourselves. In our implementation of \cite{wang2018voicefilter}, the initial SDR value (1.26dB) and the evaluated result (9.04dB) are different from that in \cite{wang2018voicefilter}, which are 10.1dB and 17.9dB respectively. However, the improvement of SDR value (7.78dB) in our implementation is the same as in \cite{wang2018voicefilter} (7.8dB). As shown in Table 3, VoiceFilter \cite{wang2018voicefilter} is the baseline that we implemented, and the Atss-Net is the model we proposed. Specifically, the SDR value that our model improved is 9.30dB, while the baseline is 7.78dB. Therefore, our proposed model outperforms the baseline method. Moreover, the parameters of our model (4.68M) are nearly 50\% less than the baseline (9.45M).

Furthermore, compared with using attention blocks (9.30dB), the SDR gain of Atss-Net without them (3.88dB) is relatively low, which indicates the effectiveness of the attention mechanism used here. Meanwhile, we compare our model with the model using the permutation invariant training (PIT) loss \cite{yu2017permutation}. This model has the same architecture as the Atss-Net, except that there is no speaker embedding as input. In this case, this PIT architecture needs to know the exact number of speakers in advance, and chooses the output that is the closest to the ground truth, i.e., with the lowest loss value. As shown in Table 3, using PIT loss to train the model (8.57dB) is not as good as directly using the target speaker embedding too.

\begin{table}
\centering
\caption{System performance in the speech enhancement experiment on AISHELL-2 dataset.}
\begin{tabular}{lllll}
\toprule
\textbf{Evaluation Metric} & \multicolumn{4}{l}{~~~~~~~~~~~~~~~~~~\textbf{PESQ}}  \\
\midrule
Test SNR          & 5    & 10   & 15   & 20            \\
\midrule
Origin       & 1.23  & 1.45  & 1.84  & 2.45           \\
\midrule
VoiceFilter~\cite{wang2018voicefilter}       & 1.83 & 2.13 & 2.41 & 2.62          \\
\midrule
\textbf{Atss-Net}          & \textbf{2.06} & \textbf{2.42} & \textbf{2.87} & \textbf{3.22}          \\
\bottomrule
\end{tabular}
\end{table}

\begin{table}[tb]
\setlength{\tabcolsep}{1.7mm}{
\centering
\caption{System performance in the target speaker separation experiment on Librispeech dataset.}
\begin{tabular}{lllll}
\toprule
\multirow{2}{*}{\textbf{Model}} & \multirow{2}{*}{\textbf{\# Param}} & \multicolumn{3}{l}{~~~~~~~~~~~\textbf{Mean SDR}}                 \\
\cline{3-5}
                                &                                    & Before                & After          & Improved  \\
\midrule
VoiceFilter~\cite{wang2018voicefilter}                     & 9.45M                             & 1.26 & 9.04           & 7.78         \\
\midrule
Atss-Net (w/o Att)           & 0.65M                             &  1.26                     & 5.14 & 3.88        \\
Atss-Net (PIT)           & 4.68M                             &  1.26                     & 9.83 & 8.57        \\
\midrule
\textbf{Atss-Net}           & 4.68M                             &  1.26                     & \textbf{10.56} & \textbf{9.30}        \\
\bottomrule
\end{tabular}}
\end{table}

\vspace{-0.2cm}
\subsection{The Speech Enhancement Experiment}
Table 2 summarizes the PESQ\footnote{\url{https://github.com/ludlows/python-pesq}} performance of our method under different SNR conditions for speech enhancement. Besides, some listening samples are available online\footnote{\url{https://tinglok.netlify.com/files/interspeech2020/}}. From the samples we can see that our Atss-Net can achieve good speech enhancement performance, even there is a singer's voice and music in the background. This supports our claim that adding the target speaker embedding as an auxiliary input to the separation network can effectively extract the target speaker's voice from various kinds of vocal background noises for speech enhancement.

\vspace{-0.5cm}
\section{Conclusions}
In this paper, we propose an Attention-based neural network for target speaker separation, namely the Atss-Net, which achieves better SDR performance compared with the baseline VoiceFilter. Besides, we do an experiment to prove that adding speaker embedding as an auxiliary input can potentially gain good speech enhancement performance with vocal background noises. For future works, we plan to model the task directly in the time-domain, so that the difference of the phase between the mixed and clean utterance will not be mismatched.

\vspace{-0.2cm}
\section{Acknowledgements}

This research is funded in part by the National Natural Science Foundation of China (61773413), Key Research and Development Program of Jiangsu Province (BE2019054), Six talent peaks project in Jiangsu Province (JY-074), Science and Technology Program of Guangzhou, China (202007030011, 201903010040).

\bibliographystyle{IEEEtran}

\bibliography{mybib}

\end{document}